\documentstyle[12pt]{article}

\newcommand {\e} {\mbox{\rm e}}

\newcounter{eq}
\newcounter{sc}

\def\overleftrightarrow#1{\vbox{\ialign{##\crcr
 $\leftrightarrow$\crcr\noalign{\kern-1pt\nointerlineskip}
 $\hfil\displaystyle{#1}\hfil$\crcr}}}

\newlength{\minitwocolumn}
\begin{document}

\begin{flushright}
DPUR/TH/52\\
September, 2016\\
\end{flushright}
\vspace{20pt}

\pagestyle{empty}
\baselineskip15pt

\begin{center}
{\large\bf Cosmology in Weyl Transverse Gravity
\vskip 1mm }

\vspace{20mm}
Ichiro Oda \footnote{E-mail address:\ ioda@phys.u-ryukyu.ac.jp
}

\vspace{5mm}
           Department of Physics, Faculty of Science, University of the 
           Ryukyus,\\
           Nishihara, Okinawa 903-0213, Japan.\\

\end{center}


\vspace{5mm}
\begin{abstract}
We study the Friedmann-Lemaitre-Robertson-Walker (FLRW) cosmology in the Weyl-transverse (WTDiff) gravity
in a general space-time dimension. The WTDiff gravity is invariant under both the local Weyl (conformal) 
transformation and the volume preserving diffeormorphisms (transverse diffeomorphisms) and is believed to 
be equivalent to general relativity at least at the classical level (perhaps, even in the quantum regime). 
It is explicitly shown by solving the equations of motion that the FLRW metric is a classical solution 
in the WTDiff gravity only when the spatial metric is flat, that is, the Euclidean space, 
and the lapse function is a nontrivial function of the scale factor. 
\end{abstract}

\newpage
\pagestyle{plain}
\pagenumbering{arabic}


\rm
\section{Introduction}

In a series of recent papers \cite{Oda1}-\cite{Oda3}, we have investigated some classical implications 
of a gravitational model called the Weyl-transverse (WTDiff) gravity \cite{Izawa}-\cite{Alvarez4}, 
which is invariant under both the local Weyl (conformal) transformation and a restricted subgroup 
of general coordinate transformation or diffeomorphisms (Diff), that is, the volume preserving 
diffeormorphisms or transverse diffeomorphisms (TDiff). We have already demonstrated that 
general relativity and the WTDiff gravity are obtained via gauge fixing procedure for 
a different local symmetry from the conformally invariant scalar-tensor gravity \footnote{The conformally invariant 
gravity theory has a wide application in phenomenology and cosmology \cite{Oda4}-\cite{Oda7}.}, 
which is a more underlying model in the sense that the conformally invariant scalar-tensor gravity 
is invariant under both the Weyl transformation and the full group of Diff. Indeed, general relativity 
is obtained by gauge-fixing the Weyl symmetry while the WTDiff gravity is reached by doing the longitudinal 
diffeomorphism from the conformally invariant scalar-tensor gravity.

This relation between general relativity and the WTDiff gravity gives us an interesting observation
that the two theories are equivalent to each other although they have different local symmetries
so that they belong to different universality classes. In fact, we have shown that in the WTDiff gravity 
the Schwarzschild metric and the Reissner-Nordstrom metric in the Cartesian coordinate system are 
classical solutions to the equations of motion as in general relativity, but they are not so in the
other coordinate systems like the spherical coordinate one. This peculiar dependence of classical 
solutions on the coordinate systems in the WTDiff gravity stems from the fact that
the TDiff are defined as a subgroup of the full Diff in such a way that the determinant of the
transformation matrix is the unity
\begin{eqnarray}
J \equiv \det J^\alpha_{\mu \prime} \equiv \det \frac{\partial x^\alpha}{\partial x^{\mu \prime}} = 1.
\label{TDiff}
\end{eqnarray}
When we transform a metric from the Cartesian coordinate system to the spherical one,
we encounter the non-trivial Jacobian factor in four dimensions
\begin{eqnarray}
J = r^2 \sin \theta,
\label{J}
\end{eqnarray}
by which the Schwarzschild and Reissner-Nordstrom metrics are not classical solutions in the spherical 
coordinate system in the WTDiff gravity.

The equivalence between general relativity and the WTDiff gravity might make it possible to tackle some
difficult problems existing in general relativity within the framework of the WTDiff gravity. For instance,
the Weyl symmetry in the WTDiff gravity can be viewed as a $\it{fake}$ Weyl symmetry, \footnote{The Weyl symmetry
in the conformally-invariant scalar-tensor gravity as well as the WTDiff gravity is sometimes called a $\it{fake}$ 
Weyl symmetry \cite{Jackiw} since this Weyl symmetry appears as a local symmetry  
in an action whenever one replaces the metric tensor $g_{\mu\nu}$ with a Weyl-invariant metric tensor 
$\hat g_{\mu\nu}$ as seen later in Eq. (\ref{Weyl-inv objects}). 
For instance, the conformally-invariant scalar-tensor gravity is obtained by replacing $g_{\mu\nu}$ in the
Einstein-Hilbert action by $\hat g_{\mu\nu}$ (In this sense, $\varphi$ is called a spurion field.) The "fakeness"
is mathematically reflected in the fact that the Noether current for this Weyl symmetry is identically vanishing
\cite{Jackiw, Oda1}.} 
which might play an important role in the cosmological constant problem. Namely, at the classical level, the $\it{fake}$ 
Weyl symmetry forbids operators of dimension zero such as the cosmological constant in the action. Then, we expect 
that the $\it{fake}$ Weyl symmetry does not give rise to a Weyl anomaly at the quantum level owing to its "fakeness"
\cite{Oda1}.
In other words, the $\it{fake}$ Weyl symmetry survives at the quantum level, thereby suppressing the radiative 
corrections to the cosmological constant. If this conjecture were true, the cosmological constant problem would become 
a mere problem of how to determine the initial value of the cosmological constant.

In the present article, in order to understand more classical implications of the WTDiff gravity, we wish to 
apply the WTDiff gravity to cosmology and find what universe is realized in the WTDiff gravity by solving 
the equations of motion on the basis of Friedmann-Lemaitre-Robertson-Walker (FLRW) metric. To do that, 
one needs to introduce a real scalar field as the matter field which must be invariant under 
the Weyl transformation and TDiff at the same time. 

This paper is organised as follows: In Section 2, we couple a real scalar field as the matter field to the WTDiff gravity 
in a consistent way, and examine the energy-momentum tensor and its conservation law. In Section 3, we show that 
the FLRW metric with the spatial flat metric and a nontrivial lapse function is in fact a classical solution.
The final section is devoted to discussions.

\section{Coupling scalar field to the Weyl-transverse (WTDiff) gravity}

We will start with an action of the conformally invariant scalar-tensor gravity in a general $n$ space-time 
dimension \footnote{We follow notation and conventions by Misner et al.'s textbook \cite{MTW}, for instance, 
the flat Minkowski metric $\eta_{\mu\nu} = diag(-, +, +, +)$, the Riemann curvature tensor 
$R^\mu \, _{\nu\alpha\beta} = \partial_\alpha \Gamma^\mu_{\nu\beta} - \partial_\beta \Gamma^\mu_{\nu\alpha} 
+ \Gamma^\mu_{\sigma\alpha} \Gamma^\sigma_{\nu\beta} - \Gamma^\mu_{\sigma\beta} \Gamma^\sigma_{\nu\alpha}$, 
and the Ricci tensor $R_{\mu\nu} = R^\alpha \, _{\mu\alpha\nu}$.
The reduced Planck mass is defined as $M_p = \sqrt{\frac{c \hbar}{8 \pi G}} = 2.4 \times 10^{18} GeV$.
Throughout this article, we adopt the reduced Planck units where we set $c = \hbar = M_p = 1$.
In this units, all quantities become dimensionless. 
Finally, note that in the reduced Planck units, the Einstein-Hilbert Lagrangian density takes the form
${\cal L}_{EH} = \frac{1}{2} \sqrt{-g} R$.} 
\begin{eqnarray}
S = \int d^n x \ \sqrt{-g} \left[ \frac{n-2}{8(n-1)} \varphi^2 R +  \frac{1}{2}
g^{\mu\nu} \partial_\mu \varphi \partial_\nu \varphi  \right],
\label{Cof-inv S-T Action}
\end{eqnarray}
which is invariant under both the Weyl transformation and diffeomorphisms (Diff). The Weyl transformation
is defined for the metric tensor $g_{\mu\nu}$ and the ghost-like scalar field $\varphi$ as  
\begin{eqnarray}
g_{\mu\nu} \rightarrow g^\prime_{\mu\nu} = \Omega^2(x) g_{\mu\nu}, \quad
\varphi \rightarrow \varphi^\prime = \Omega^{- \frac{n-2}{2}}(x) \varphi,
\label{Weyl transf}
\end{eqnarray}
where $\Omega(x)$ is an arbitrary scalar function. 

From this fundamental action (\ref{Cof-inv S-T Action}), we can derive the Einstein-Hilbert action
of general relativity by taking the gauge condition 
\begin{eqnarray}
\varphi = 2 \sqrt{\frac{n-1}{n-2}},
\label{GR gauge}
\end{eqnarray}
for the Weyl symmetry. 

On the other hand, the gauge condition 
\begin{eqnarray}
\varphi = 2 \sqrt{\frac{n-1}{n-2}} |g|^{- \frac{n-2}{4n}},
\label{WTDiff gauge}
\end{eqnarray}
for the longitudinal diffeomorphism leads to an action of the WTDiff gravity \cite{Alvarez1, Alvarez2,
Alvarez3, Alvarez4, Oda2, Oda3}
\begin{eqnarray}
S_g = \frac{1}{2} \int d^n x \ |g|^{\frac{1}{n}} \left[ R + \frac{(n-1)(n-2)}{4n^2} \frac{1}{|g|^2}
g^{\mu\nu} \partial_\mu |g| \partial_\nu |g|  \right],
\label{WTDiff Action}
\end{eqnarray}
where we have defined $g = \det g_{\mu\nu} < 0$. 
Thus, the WTDiff gravity is at least classically equivalent to general relativity since 
the both actions can be derived via the different choices of gauge condition from the same 
action (\ref{Cof-inv S-T Action}). We conjecture that the equivalence would be valid even at
the quantum level since the $\it{fake}$ Weyl symmetry and the longitudinal diffeomorphism
do not seem to possess the corresponding anomalies.

Now we wish to construct the matter action of a real scalar field $\phi$ coupled to gravity in the Weyl and 
TDiff-invariant manner \cite{Oda1}. To do so, let us consider first an action of the real scalar field $\phi$ 
with the potential term $V(\phi)$
\begin{eqnarray}
S_m^{Diff} = \int d^n x \ |g|^{\frac{1}{2}} \left[ - \frac{1}{2} g^{\mu\nu} \partial_\mu \phi \partial_\nu \phi
- V(\phi) \right].
\label{Diff Matt-Action}
\end{eqnarray}
Note that $S_m^{Diff}$ is manifestly invariant under the full Diff. 

Next, to make $S_m^{Diff}$ be invariant under the Weyl transformation as well, it is necessary to construct 
the Weyl-invariant objects
\begin{eqnarray}
\hat g_{\mu\nu} = \left( \frac{1}{2} \sqrt{\frac{n-2}{n-1}} \varphi \right)^{\frac{4}{n-2}} g_{\mu\nu},
\quad \hat \phi = \frac{\phi}{\varphi},
\label{Weyl-inv objects}
\end{eqnarray}
where we have assumed that the real scalar field $\phi$ has the same transformation property under the Weyl
transformation (\ref{Weyl transf}) as the ghost-like scalar field $\varphi$. 
In order to obtain the Weyl and Diff-invariant matter action,
it is enough to replace $g_{\mu\nu}$ and $\phi$ in the action (\ref{Diff Matt-Action}) by the corresponding
Weyl-invariant objects $\hat g_{\mu\nu}$ and $\hat \phi$. The resulting action takes the form
\begin{eqnarray}
S_m^{WDiff} 
&=& \int d^n x \ |\hat g|^{\frac{1}{2}} \left[ - \frac{1}{2} \hat g^{\mu\nu} 
\partial_\mu \hat \phi \partial_\nu \hat \phi- V(\hat \phi) \right]  \nonumber\\
&=& \int d^n x \ |g|^{\frac{1}{2}} \Biggl[ - \frac{1}{2} \left( \frac{1}{2} \sqrt{\frac{n-2}{n-1}} 
\varphi \right)^2 g^{\mu\nu} \partial_\mu \left(\frac{\phi}{\varphi} \right) 
\partial_\nu \left(\frac{\phi}{\varphi} \right)    \nonumber\\
&-& \left( \frac{1}{2} \sqrt{\frac{n-2}{n-1}} 
\varphi \right)^{\frac{2n}{n-2}} V \left(\frac{\phi}{\varphi}\right) \Biggr].
\label{WDiff Matt-Action}
\end{eqnarray}

Finally, reducing further the Weyl and Diff-invariant matter action (\ref{WDiff Matt-Action}) to 
the Weyl and TDiff-invariant matter action requires us to take the gauge condition (\ref{WTDiff gauge})
for the longitudinal diffeomorphism. Consequently, we have the WTDiff-invariant matter action given by
\begin{eqnarray}
S_m^{WTDiff} 
&=& \int d^n x \ |g|^{\frac{1}{2}} \Biggl\{ - \frac{1}{8} \frac{n-2}{n-1} \ g^{\mu\nu}
\left[ \left(\frac{n-2}{4n}\right)^2 \frac{\phi^2}{|g|^2} \partial_\mu |g| \partial_\nu |g| 
+ \frac{n-2}{2n} \frac{\phi}{|g|} \partial_\mu |g| \partial_\nu \phi
+ \partial_\mu \phi \partial_\nu \phi \right]     \nonumber\\
&-& |g|^{- \frac{1}{2}} V \left(\frac{1}{2} \sqrt{\frac{n-2}{n-1}} 
|g|^{\frac{n-2}{4n}} \phi \right) \Biggr\}.
\label{WTDiff Matt-Action}
\end{eqnarray}

From this action, the equation of motion for $\phi$ is derived to be
\begin{eqnarray}
&{}& \frac{1}{8} \frac{n-2}{n-1} |g|^{\frac{1}{2}} 
\left[ \frac{(n-2)(5n-2)}{8n^2} \frac{\phi}{|g|^2} \left(\partial_\rho |g| \right)^2 
- \frac{n-2}{2n} \frac{\phi}{|g|} \nabla_\rho \nabla^\rho |g| - 2 \nabla_\rho \nabla^\rho \phi \right]
\nonumber\\
&+& \frac{1}{2} \sqrt{\frac{n-2}{n-1}} 
|g|^{\frac{n-2}{4n}} V^\prime \left( \frac{1}{2} \sqrt{\frac{n-2}{n-1}} |g|^{\frac{n-2}{4n}} \phi \right)
\nonumber\\
&=& 0,
\label{Phi-eq}
\end{eqnarray}
where  we have defined $\nabla_\mu \nabla_\nu |g| = \partial_\mu \partial_\nu |g| 
- \Gamma^\rho_{\mu\nu} \partial_\rho |g|$ and the prime on the potential $V$ denotes the differentiation 
with respect to its argument. Furthermore, taking the variation of the action (\ref{WTDiff Matt-Action}) 
with respect to the metric tensor produces an expression, which is proportional to the energy-momentum tensor 
of the scalar matter field
\begin{eqnarray}
\frac{\delta S_m^{WTDiff}}{\delta g^{\mu\nu}}
&=& - \frac{1}{8} \frac{n-2}{n-1} |g|^{\frac{1}{2}} 
\Biggl\{ \left(\frac{n-2}{4n} \right)^2 \frac{\phi^2}{|g|^2} \partial_\mu |g| \partial_\nu |g| 
+ \frac{n-2}{4n} \frac{\phi}{|g|} \left( \partial_\mu |g| \partial_\nu \phi 
+ \partial_\nu |g| \partial_\mu \phi \right) 
\nonumber\\
&+& \partial_\mu \phi \partial_\nu \phi 
+ g_{\mu\nu} \Biggl[ - \frac{5}{2} \left(\frac{n-2}{4n} \right)^2 \frac{\phi^2}{|g|^2}
\left(\partial_\rho |g| \right)^2 - \frac{n-2}{2n^2} \frac{\phi}{|g|} \partial_\rho \phi \partial^\rho |g|
- \frac{1}{n} \left( \partial_\rho \phi \right)^2 
\nonumber\\
&+& 2 \left(\frac{n-2}{4n^2} \right)^2 \frac{\phi^2}{|g|} 
\nabla_\rho \nabla^\rho |g|
+ \frac{n-2}{2n} \phi \nabla_\rho \nabla^\rho \phi \Biggr] \Biggr\}
\nonumber\\
&+& \frac{n-2}{8n} \sqrt{\frac{n-2}{n-1}} |g|^{\frac{n-2}{4n}} g_{\mu\nu} \phi \
V^\prime \left( \frac{1}{2} \sqrt{\frac{n-2}{n-1}} |g|^{\frac{n-2}{4n}} \phi \right).
\label{g-eq}
\end{eqnarray}
Then, it is convenient to rewrite this expression by eliminating the potential term
through the equation of motion for $\phi$ in Eq. (\ref{Phi-eq}) as
\begin{eqnarray}
\frac{\delta S_m^{WTDiff}}{\delta g^{\mu\nu}}
&=& - \frac{1}{8} \frac{n-2}{n-1} |g|^{\frac{1}{2}} 
\Biggl\{ \left(\frac{n-2}{4n} \right)^2 \frac{\phi^2}{|g|^2} \partial_\mu |g| \partial_\nu |g| 
+ \frac{n-2}{4n} \frac{\phi}{|g|} \left( \partial_\mu |g| \partial_\nu \phi 
+ \partial_\nu |g| \partial_\mu \phi \right) 
\nonumber\\
&+& \partial_\mu \phi \partial_\nu \phi 
+ g_{\mu\nu} \Biggl[ - \frac{1}{n} \left(\frac{n-2}{4n} \right)^2 \frac{\phi^2}{|g|^2}
\left(\partial_\rho |g| \right)^2 - \frac{n-2}{2n^2} \frac{\phi}{|g|} \partial_\rho \phi \partial^\rho |g|
- \frac{1}{n} \left( \partial_\rho \phi \right)^2 \Biggr] \Biggr\}
\nonumber\\
&\equiv& - \frac{1}{2} |g|^{\frac{1}{2}} \left( T_{(m) \mu\nu} 
- \frac{1}{n} g_{\mu\nu} T_{(m)} \right),
\label{g-eq 2}
\end{eqnarray}
where $T_{(m) \mu\nu}$ is defined as 
\begin{eqnarray}
T_{(m) \mu\nu}
&=& \frac{1}{4} \frac{n-2}{n-1} 
\Biggl[ \left(\frac{n-2}{4n} \right)^2 \frac{\phi^2}{|g|^2} \partial_\mu |g| \partial_\nu |g| 
+ \frac{n-2}{4n} \frac{\phi}{|g|} \left( \partial_\mu |g| \partial_\nu \phi 
+ \partial_\nu |g| \partial_\mu \phi \right)
\nonumber\\ 
&+& \partial_\mu \phi \partial_\nu \phi \Biggr].
\label{T_m}
\end{eqnarray}

Without the matter action of the scalar field, from the action (\ref{WTDiff Action}) of the WTDiff gravity, 
the Einstein equations, which are obtained by taking the variation with respect to the metric tensor, 
take the form
\begin{eqnarray}
R_{\mu\nu} - \frac{1}{n} g_{\mu\nu} R = T_{(g) \mu\nu} - \frac{1}{n} g_{\mu\nu} T_{(g)},
\label{Eq of motion of WTDiff}
\end{eqnarray}
where $T_{(g) \mu\nu}$ is defined by
\begin{eqnarray}
T_{(g) \mu\nu} = \frac{(n-2)(2n-1)}{4n^2} \frac{1}{|g|^2} \partial_\mu |g| \partial_\nu |g|
- \frac{n-2}{2n} \frac{1}{|g|} \nabla_\mu \nabla_\nu |g|.
\label{T_g}
\end{eqnarray}
Since we regard the sum of the action (\ref{WTDiff Action}) of the WTDiff gravity plus that
(\ref{WDiff Matt-Action}) of the scalar matter field as a total action, the Einstein equations
at hand are given by
\begin{eqnarray}
R_{\mu\nu} - \frac{1}{n} g_{\mu\nu} R = T_{\mu\nu} - \frac{1}{n} g_{\mu\nu} T,
\label{Einstein Eq}
\end{eqnarray}
where the total energy-momentum tensor $T_{\mu\nu}$ is now defined by
\begin{eqnarray}
T_{\mu\nu} &=& T_{(g) \mu\nu} + T_{(m) \mu\nu}
\nonumber\\
&=& \frac{(n-2)(2n-1)}{4n^2} \frac{1}{|g|^2} \partial_\mu |g| \partial_\nu |g|
- \frac{n-2}{2n} \frac{1}{|g|} \nabla_\mu \nabla_\nu |g|
\nonumber\\
&+& \frac{1}{4} \frac{n-2}{n-1} 
\Biggl[ \left(\frac{n-2}{4n} \right)^2 \frac{\phi^2}{|g|^2} \partial_\mu |g| \partial_\nu |g| 
+ \frac{n-2}{4n} \frac{\phi}{|g|} \left( \partial_\mu |g| \partial_\nu \phi 
+ \partial_\nu |g| \partial_\mu \phi \right)
\nonumber\\ 
&+& \partial_\mu \phi \partial_\nu \phi \Biggr].
\label{T}
\end{eqnarray}
Let us note that the Einstein equations of the WTDiff gravity with the WTDiff-invariant
matter field have the $\it{traceless}$ form, which is also a common feature to unimodular gravity
\cite{Einstein}-\cite{Oda10}.

\section{Cosmology}
 
Now we are ready to turn our attention to cosmology in the WTDiff gravity with the scalar matter
whose equations of motion are of form (\ref{Einstein Eq}). Before attempting to solve the Einstein 
equations (\ref{Einstein Eq}), let us notice that as given in Eq. (\ref{T}), the energy-momentum tensor 
has a rather complicated structure owing to the presence of the determinant of the metric tensor, 
which makes it difficult to solve the Einstein equations analytically. Thus, we will select the gauge 
condition 
\begin{eqnarray}
g = -1,
\label{g=-1}
\end{eqnarray}
for the Weyl symmetry. This choice of the gauge condition provides us with an enormous simplication 
since the energy-momentum tensor (\ref{T}) is reduced to the tractable form
\begin{eqnarray}
T_{\mu\nu} = \frac{1}{4} \frac{n-2}{n-1} \partial_\mu \phi \partial_\nu \phi.
\label{T in g=-1}
\end{eqnarray}

It is usually assumed that our universe is described in terms of an expanding, homogeneous and isotropic
Friedmann-Lemaitre-Robertson-Walker (FLRW) universe given by the line element 
\begin{eqnarray}
d s^2 &=& g_{\mu\nu} d x^\mu d x^\nu    \nonumber\\
&=& - d t^2 + a^2(t) \gamma_{ij}(x) d x^i d x^j,
\label{Line element 1}
\end{eqnarray}
where $a(t)$ is a scale factor and $\gamma_{ij}(x)$ is the spatial metric of the unit $(n-1)$-sphere,
unit $(n-1)$-hyperboloid or $(n-1)$-plane, and $i, j$ run over spatial coordinates ($i = 1, 2, \cdots, n-1$).
However, this metric ansatz does not satisfy the gauge condition (\ref{g=-1}) so the line element (\ref{Line element 1})
should be somewhat modified. A suitable modification, which respects the gauge condition (\ref{g=-1}),
is to work with the following line element;
\begin{eqnarray}
d s^2 = - N^2(t) d t^2 + a^2(t) (d x^i)^2,
\label{Line element 2}
\end{eqnarray}
where $N(t)$ is a lapse function and the spatial geometry is chosen to be the $(n-1)$-plane, i.e., the $(n-1)$-dimensional 
Euclidean space. Note that the existence of the lapse function $N(t)$ means that a time coordinate $t$ does not 
coincide with the proper time of particles at rest.
With this line element, the gauge condition (\ref{g=-1}) provides 
a relation between the lapse function $N(t)$ and the scale factor  $a(t)$
\begin{eqnarray}
N(t) = a^{-(n-1)}(t).
\label{N vs a}
\end{eqnarray}

Given the line element (\ref{Line element 2}) and Eq. (\ref{N vs a}), it turns out 
that the non-vanishing components of the $\it{traceless}$ Einstein tensor, which is defined as
\begin{eqnarray}
G^T_{\mu\nu} = R_{\mu\nu} - \frac{1}{n} g_{\mu\nu} R,
\label{traceless Eins-tensor}
\end{eqnarray}
are given by
\begin{eqnarray}
G^T_{tt} &=& - \frac{(n-1)(n-2)}{n} \left[ \dot{H}  + (n-1) H^2 \right],  \nonumber\\
G^T_{ij} &=& - \frac{n-2}{n} a^{2n} \left[ \dot{H}  + (n-1) H^2 \right] \delta_{ij},
\label{traceless Eins-tensor 2}
\end{eqnarray}
where $H = \frac{\dot{a}}{a}$ is the Hubble parameter and we have defined $\dot{a} = \frac{d a(t)}{dt}$. 
In a similar way, the non-vanishing components 
of the $\it{traceless}$ energy-momentum tensor, which is defined as
\begin{eqnarray}
T^T_{\mu\nu} = T_{\mu\nu} - \frac{1}{n} g_{\mu\nu} T,
\label{traceless stress-tensor}
\end{eqnarray}
read
\begin{eqnarray}
T^T_{tt} &=& \frac{n-2}{4n} (\dot{\phi})^2,  \nonumber\\
T^T_{ij} &=& \frac{1}{n-1} \frac{n-2}{4n} a^{2n} (\dot{\phi})^2 \delta_{ij},
\label{traceless stress-tensor 2}
\end{eqnarray}
where we have specified the scalar field $\phi$ to be spatially homogeneous, that is, $\phi = \phi(t)$.
As a result, the $\it{traceless}$ Einstein equations (\ref{Einstein Eq}) are cast to
be a single equation
\begin{eqnarray}
\dot{H} + (n-1) H^2 = - \frac{1}{4(n-1)} (\dot{\phi})^2.
\label{Single eq}
\end{eqnarray}
Moreover, using the line element (\ref{Line element 2}) and Eq. (\ref{N vs a}), the equation of motion 
for the scalar field $\phi$, Eq. (\ref{Phi-eq}), is simplified to be
\begin{eqnarray}
\ddot{\phi} + 2(n-1) H \dot{\phi} + 2 \sqrt{\frac{n-1}{n-2}} a^{-2(n-1)}
V^\prime \left( \frac{1}{2} \sqrt{\frac{n-2}{n-1}} \phi \right) = 0.
\label{Phi-eq 2}
\end{eqnarray}

It is of interest to see that the $\it{traceless}$ Einstein equations (\ref{Einstein Eq}) have yielded only
the single equation (\ref{Single eq}), which is similar to the Raychaudhuri equation or the first Friedmann 
equation \cite{Rubakov, Mukhanov} that comes from all $ij$-components of the Einstein equations in general relativity 
though there is a slight difference in Eq. (\ref{Single eq}) which will be mentioned shortly. However, 
in the present formalism, the (second) Friedmann equation stemming from $00$-component of the Einstein equations is missing. 
In order to solve Eq. (\ref{Single eq}), we need the conservation law of the energy-momentum tensor.
In this respect, recall that in general relativity the first Friedmann equation can be viewed as a consequence
of the (second) Friedmann equation and covariant conservation of energy, so that the combination of 
the (second) Friedmann equation and the conservation law, supplemented by the equation of state
$p = p(\rho)$ (which will appear later), forms a complete system of equations that determines the
two unknown functions, the scale factor $a(t)$ and energy density $\rho$. In our formalism, instead of
the (second) Friedmann equation, we have to use the first Friedmann equation like Eq. (\ref{Single eq}).   

At this stage, it is worth stressing that the energy-momentum tensor (\ref{T}) is not covariantly conserved
since it is derived from the action which is not invariant under the general coordinate transformation
(Diff) but only invariant under the Weyl transformation and TDiff. Actually, the following (well-known) proof
clarifies the reason why the energy-momentum tensor constructed out of a Diff-invariant action is only covariantly
conserved: Suppose that a generic action $S$ is invariant under Diff 
\begin{eqnarray}
S = \int d^n x \sqrt{-g} \cal{L}.
\label{S}
\end{eqnarray}
Under Diff, the metric tensor transforms as
\begin{eqnarray}
\delta g^{\mu\nu} = \nabla^\mu \xi^\nu + \nabla^\nu \xi^\mu, 
\label{GCT}
\end{eqnarray}
where $\xi^\mu$ is a local parameter of Diff. Under Diff, the action $S$ is transformed into
\begin{eqnarray}
\delta S = - \int d^n x \sqrt{-g} T_{\mu\nu} \nabla^\mu \xi^\nu, 
\label{delta S}
\end{eqnarray}
where the energy-momentum tensor $T_{\mu\nu}$ is defined as
\begin{eqnarray}
T_{\mu\nu} &=& - \frac{2}{\sqrt{-g}} \frac{\delta (\sqrt{-g} \cal{L})}{\delta g^{\mu\nu}}
\nonumber\\
&=& -2 \frac{\delta \cal{L}}{\delta g^{\mu\nu}} + g_{\mu\nu} \cal{L}. 
\label{def T}
\end{eqnarray}
By integrating by parts, Eq. (\ref{delta S}) can be recast to the form
\begin{eqnarray}
\delta S = \int d^n x \sqrt{-g} \nabla_\mu T^{\mu\nu} \xi_\nu, 
\label{delta S 2}
\end{eqnarray}
from which we can arrive at the covariant conservation law of the energy-momentum tensor
\begin{eqnarray}
\nabla_\mu T^{\mu\nu} = 0. 
\label{Conserv-law}
\end{eqnarray}
Let us note that the general coordinate invariance of the action plays a critical role 
in this proof. 

Accordingly, in order to derive the energy-momentum tensor satisfying the covariant 
conservation law, we must make use of not $S_m^{WTDiff}$ in (\ref{WTDiff Matt-Action}) but 
$S_m^{WDiff}$ in (\ref{WDiff Matt-Action}). \footnote{The contribution from $S$ in 
(\ref{Cof-inv S-T Action}) vanishes in the gauge condition (\ref{g=-1}).} 
After a straightforward calculation, using the gauge condition (\ref{g=-1}), 
the energy-momentum tensor reads
\begin{eqnarray}
T^{(cov)}_{\mu\nu} = \frac{1}{4} \frac{n-2}{n-1} \partial_\mu \phi \partial_\nu \phi
+ g_{\mu\nu} \left[ - \frac{1}{8} \frac{n-2}{n-1} (\partial_\rho \phi)^2 
- V \left( \frac{1}{2} \sqrt{\frac{n-2}{n-1}} \phi \right) \right].
\label{Cov T in g=-1}
\end{eqnarray}
In contrast to the previous result (\ref{T in g=-1}), in this case the terms proportional to
$g_{\mu\nu}$ have emerged, by which the energy-momentum tensor (\ref{Cov T in g=-1}) turns out
to be covariantly conserved by using the equation of motion for $\phi$ in Eq. (\ref{Phi-eq})
together with the gauge condition (\ref{g=-1}). The non-vanishing components of $T^{(cov) \mu} \, _\nu$ 
are easily evaluated to be 
\begin{eqnarray}
T^{(cov) t} \, _t &=& - \frac{1}{8} \frac{n-2}{n-1} a^{2(n-1)} (\dot{\phi})^2 
- V \left( \frac{1}{2} \sqrt{\frac{n-2}{n-1}} \phi \right) \equiv - \rho(t),
\nonumber\\
T^{(cov) i} \, _j &=& \left[ \frac{1}{8} \frac{n-2}{n-1} a^{2(n-1)} (\dot{\phi})^2 
- V \left( \frac{1}{2} \sqrt{\frac{n-2}{n-1}} \phi \right) \right] \delta^i \, _j
\equiv p(t) \delta^i \, _j,
\label{Comp-T}
\end{eqnarray}
where we have introduced energy density $\rho(t)$ and pressure $p(t)$ in a conventional way.
Then, the covariant conservation law (\ref{Conserv-law}) leads to an equation
\begin{eqnarray}
\dot{\rho} + (n-1) H ( \rho + p ) = 0.
\label{rho-p}
\end{eqnarray}

To close the system of equations, which determine the dynamics of homogeneous and isotropic universe,
we have to specify the equation of state of matter as usual
\begin{eqnarray}
p = w \rho,
\label{w}
\end{eqnarray}
where $w$ is a certain constant. Of course, the equation of state is not a consequence of equations of 
our formalism, but should be determined by matter content in our universe. With the help of Eq. (\ref{w}),
Eq. (\ref{rho-p}) is exactly solved to be
\begin{eqnarray}
\rho(t) = \rho_0 a^{-(n-1)(w+1)}(t),
\label{rho-a}
\end{eqnarray}
where $\rho_0$ is an integration constant. Eqs. (\ref{rho-p})-(\ref{rho-a}) are the same expressions as
in general relativity.
Now, using Eqs. (\ref{Comp-T}), (\ref{w}) and (\ref{rho-a}), our Friedmann equation (\ref{Single eq})
is rewritten as
\begin{eqnarray}
\dot{H} + (n-1) H^2 = - \frac{w+1}{n-2} \rho_0 a^{-(n-1)(w+3)}.
\label{Friedmann eq}
\end{eqnarray}

Since it is difficult to find a general solution to this equation (\ref{Friedmann eq}), we will refer to only
special solutions which are physically interesting. Looking at the RHS in Eq. (\ref{Friedmann eq}), one
soon notices that at $w=-1$ and $w=-3$, specific situations occur. Actually, at $w=-1$, Eq. (\ref{Friedmann eq})
can be exactly integrated to be 
\begin{eqnarray}
a(t) = a_0 t^{\frac{1}{n-1}},
\label{a at w=-1}
\end{eqnarray}
where $a_0$ is an integration constant and this solution describes the decelerating universe in four dimensions
owing to $\ddot{a} <0$.

At the case $w=-3$, Eq. (\ref{Friedmann eq}) is reduced to the form
\begin{eqnarray}
\dot{H} + (n-1) H^2 = \frac{2}{n-2} \rho_0.
\label{Friedmann eq at w=-3}
\end{eqnarray}
This equation includes a special solution describing an exponentially expanding universe 
\begin{eqnarray}
a(t) = a_0 \e^{H_0 t},
\label{a at w=-3}
\end{eqnarray}
where $H_0$ is a constant defined as
\begin{eqnarray}
H_0 = \sqrt{\frac{2 \rho_0}{(n-1)(n-2)}}.
\label{H_0}
\end{eqnarray}

Finally, one can find a special solution such that the scale factor $a(t)$ is the form of polynomial 
in $t$ 
\begin{eqnarray}
a(t) = a_0 t^\alpha,
\label{Poly a}
\end{eqnarray}
where $\alpha$ is a constant to be determined by the Friedmann equation (\ref{Friedmann eq}). 
It is easy to verify that the constant $\alpha$ is given by
\begin{eqnarray}
\alpha = \frac{2}{(n-1)(w+3)},
\label{alpha}
\end{eqnarray}
so that in this case the scale factor takes the form
\begin{eqnarray}
a(t) = a_0 t^{\frac{2}{(n-1)(w+3)}},
\label{Scale factor}
\end{eqnarray}
which includes the solution (\ref{a at w=-1}) when $w = -1$.
Then, the accelerating universe $\ddot{a}(t) > 0$ requires 
\begin{eqnarray}
w < \frac{-3n+5}{n-1},
\label{accelerating univ}
\end{eqnarray}
while the decelerating universe does
\begin{eqnarray}
w > \frac{-3n+5}{n-1}.
\label{decelerating univ}
\end{eqnarray}

One might wonder how the obtained solutions are related to solutions in general relativity.
In particular, in general relativity we are familiar with the fact that the case $w=-1$ corresponds 
to the cosmological constant and the solution is then an exponentially expanding universe whereas
in our case the corresponding solution belongs to the case $w=-3$, which appears to be strange. But this is 
just an illusion since we do not use the conventional form (\ref{Line element 1}) of the line element 
but the line element (\ref{Line element 2}) involving the nontrivial lapse function $N(t)$.

In order to show that our result coincides with that in general relativity, let us focus our
attention to the Friedmann equation (\ref{Single eq}). By means of Eq. (\ref{Comp-T}), this equation
is rewritten as    
\begin{eqnarray}
\dot{H} + (n-1) H^2 = - \frac{1}{n-2} N^2 (\rho + p),
\label{Single eq 2}
\end{eqnarray}
where we recovered the lapse function $N(t)$ by using Eq. (\ref{N vs a}).
  
On the other hand, with the conventional notation of the energy-momentum tensor
\begin{eqnarray}
T^\mu \, _\nu = diag ( -\rho, p, \cdots, p ),
\label{Cov-T}
\end{eqnarray}
and the line element (\ref{Line element 2}), the Einstein equations in general relativity 
\begin{eqnarray}
G^\mu \, _\nu \equiv R^\mu \, _\nu - \frac{1}{2} \delta^\mu \, _\nu R = T^\mu \, _\nu,
\label{Cov-Ein-eq}
\end{eqnarray}
become a set of the Friedmann equations   
\begin{eqnarray}
&{}& H^2 = \frac{2}{(n-1)(n-2)} N^2 \rho,  
\label{Cov-Fried-eq 1} \\
&{}& \dot{H} + \frac{n-1}{2} H^2 - \frac{\dot{N}}{N} H = - \frac{1}{n-2} N^2 p.
\label{Cov-Fried-eq 2}
\end{eqnarray}
By using Eq. (\ref{N vs a}), Eq. (\ref{Cov-Fried-eq 2}) is written as 
\begin{eqnarray}
\dot{H} + \frac{3(n-1)}{2} H^2 = - \frac{1}{n-2} N^2 p.
\label{Cov-Fried-eq 2-1}
\end{eqnarray}
Eq. (\ref{Cov-Fried-eq 1}) allows us to rewrite this equation to the form
\begin{eqnarray}
\dot{H} + (n-1) H^2 = - \frac{1}{n-2} N^2 (\rho + p),
\label{Cov-Fried-eq 2-2}
\end{eqnarray}
which coincides with our Friedmann equation (\ref{Single eq 2}). This demonstration
clearly indicates that our cosmological solution is just equivalent to that of
general relativity specified in such a way that the line element is (\ref{Line element 2})
and the lapse function is given by Eq. (\ref{N vs a}).

\section{Discussions}

In this article, we have studied the Friedmann-Lemaitre-Robertson-Walker (FLRW) cosmology 
in the framework of the Weyl-transverse (WTDiff) gravity in a general space-time dimension.
One of interesting aspects of our result is that spatial geometry is completely selected
to be a flat Euclidean space among three possibilities, those are, the unit sphere, unit
hyperboloid and Euclidean space. In regards to this, let us recall that for both closed and 
open universes, the spatial curvature can be often be neglected, so one can use the spatially
flat metric and this is certainly possible for processes happened at spatial scales much smaller
than the curvature radius $a(t)$. Our result insists that the spatial metric must be flat 
at least in the classical level where the present analysis could be applied.    

Furthermore, our result requires that the lapse function, which is usually taken to be 1 
by hand, should be a nontrivial function of the scale factor. Note that there is $\it{a \ priori}$ 
no need for fixing the lapse function to be a certain value. To put differently, there is no need
for choosing a time coordinate such that it agrees with proper time of particles at rest
since world lines of particles at rest are geodesic even in the line element with the
nontrivial lapse function of time coordinate $t$.

As future problems, we would like to list up two different problems. One problem is to
look for a broad class of classical solutions which do not satisfy the gauge condition (\ref{g=-1}).
As seen in  (\ref{T}), the total energy-momentum tensor involves the complicated contribution
from the metric determinant and this contribution behaves as if it were the source of a new matter field 
in the traceless Einstein equations. This fact makes it quite difficult to find classical solutions
except the case $g = -1$, or more generally, $g =$ constant.

Another interesting and important problem is to investigate quantum aspect of the present formalism.
The most attractive point in the present formalism is the existence of the fake Weyl symmetry,
by which the cosmological constant cannot appear in the classical action. It is widely believed 
that the Weyl symmetry is violated by radiative corrections, thereby giving rise to a nonvanishing
value of the cosmological constant at the quantum level. However, we conjecture that the fake Weyl symmetry is kept 
even in the quantum regime owing to its fakeness. Our conjecture seems to be consistent with
the fact that the fake Weyl symmetry has an indentically vanishing Noether current. 
We wish to consider these problems in near future.

\begin{flushleft}
{\bf Acknowledgements}
\end{flushleft}
This work is supported in part by the Grant-in-Aid for Scientific 
Research (C) No. 16K05327 from the Japan Ministry of Education, Culture, 
Sports, Science and Technology.


\end{document}